\documentstyle[12pt,epsf]{article}

\begin{document}
\sloppy
\sloppy
\sloppy
$\ $
\begin{flushright}{UT-740,1996:\ \ hep-th/9602080
}\end{flushright}
\vskip 0.5 truecm

\begin{center}
{\large{\bf  Path integral of the hydrogen atom,
the Jacobi's principle of least action and one-dimensional quantum gravity }}
\end{center}
\vskip .5 truecm
\centerline{\bf Kazuo Fujikawa}
\vskip .4 truecm
\centerline {\it Department of Physics,University of Tokyo}
\centerline {\it Bunkyo-ku,Tokyo 113,Japan}
\vskip 0.5 truecm

\makeatletter
\@addtoreset{equation}{section}
\def\theequation{\thesection.\arabic{equation}}
\makeatother

\vskip 0.5 truecm

\begin{abstract}
A path integral 
evaluation of the Green's function for the hydrogen atom initiated by Duru and Kleinert is  studied by recognizing it  as a 
special case of the 
general treatment of the separable Hamiltonian of Liouville-type. The 
basic dynamical principle involved is identified as the Jacobi's principle of least action for given energy which is reparametrization invariant, and thus the appearance of a gauge freedom is naturally understood. 
The separation of variables in operator formalism corresponds to a choice of
gauge in path integral, and the Green's function is shown to be gauge 
independent if the operator ordering is properly taken into account. Unlike the conventional Feynman path integral,which deals with a space-time picture of particle motion, the path integral 
on the basis of the Jacobi's principle sums over orbits in space.
We illustrate these properties 
 by evaluating an  exact path integral of the Green's function for the hydrogen atom 
in parabolic coordinates, and 
thus avoiding the use of  the Kustaanheimo-Stiefel transformation. 
In the present formulation 
, the  Hamiltonian for Stark effect is converted to the one for anharmonic oscillators with an unstable quartic coupling. We also study  the hydrogen atom path integral from a view point of one-dimensional quantum gravity coupled to matter fields representing the electron coordinates. A simple BRST analysis 
of the problem with an evaluation of Weyl anomaly is presented .

\par

\end{abstract}

\newpage
\section{Introduction}
The Hamiltonian of the hydrogen atom provides one of those few 
examples which can be solved exactly, and as such various interesting 
alternative methods to solve it have been proposed in the past. Among those 
methods, one may count an elegant momentum space analysis by Schwinger[1]who
exploited the O(4) symmetry hidden in the hydrogen atom Hamiltonian. The path 
integral treatment of the problem is also interesting not only for a 
methodological interest but also for a pedagogical purpose.  A comprehensive study of a semi-classical approximation of the 
path integral for the Green's function at a given energy has been performed by
Gutzwiller[2]; in particular,  he found an exact Green's 
function for negative energy in the polar coordinates of  momentum space  by a 
semi-classical approximation. It is also known that a path integral of the 
s-wave propagator ( or evolution operator ) is obtained by summing a 
perturbative series [3]. In 1979,  Duru and Kleinert[4] showed an elegant  
path integral method to evaluate the Green's function for the hydrogen atom 
exactly. Two basic 
ingredients in their method are the use of a re-scaled time variable and
the  so called  Kustaanheimo-Stiefel transformation[5]which reveals the O(4)
symmetry explicitly in the coordinate space. Many of the clarifying works of this approach have been published [6] - [15]. The main issue in these works is a 
physical meaning of the ``re-scaled time variable''. We here study this issue 
from a completely different view point on the basis of the Jacobi's principle 
of least action by using a  general gauge theoretical technique.

We first recognize the procedure in Ref.[4] as a 
special case of the general treatment of classically separable 
Hamiltonian of Liouville-type. The basic dynamical principle involved is 
then identified as the Jacobi's principle of least action for  given energy. 
The fundamental feature of the conventional Feynman path integral, 
which is associated with  the Hamilton's principle of stationary action, is that it deals with 
a {\em space-time} picture of particle motion. On the other hand, the path integral on the basis of the Jacobi's principle of least action is basically static and analogous to geometrical optics. A space-time picture is thus lost, and one deals with a sum over orbits in {\em space} instead of space-time.   
Another characteristics of the Jacobi's
principle of least action is  that it is reparametrization invariant, and the 
appearance of a gauge freedom to fix an arbitrary   parameter, which describes 
orbits  for  fixed energy, is clearly seen. The general technique of gauge 
theory is thus applicable to the evaluation of path integral, and  
a suitable choice of gauge 
 simplifies the problem such as the hydrogen atom. In particular, the Green's 
function is  shown to be gauge independent.

The use of  the Kustaanheimo-Stiefel
transformation is  rather technical and it is not essential in solving the problem exactly. We in fact 
show a simple trick in parabolic coordinates which solves the hydrogen
atom exactly. This trick was used before in a different context by Ravndal and Toyoda[16].
We also note that the path integral in holomorphic 
variables  (or the coherent-state path integral) is convenient to evaluate
the path integral of the hydrogen atom.

As an application of the present approach, the Hamiltonian of Stark effect is shown to be reduced to anharmonic oscillators with an unstable quartic coupling, for which a resummation technique of a perturbative series is well-known. 
We also study the hydrogen atom path integral from a view point of 
one-dimensional quantum gravity, and a simple BRST analysis of the problem 
with an evaluation of the Weyl anomaly is presented. This provides an alternative way to see the gauge independence of  Green's functions. 

\section{Separable Hamiltonian of Liouville-type}
We here explain the basic procedure to treat a general separable Hamiltonian 
of Liouville-type.
We start with a separable Hamiltonian 
\begin{equation}
H= \frac{1}{V_{1}(q_{1}) + V_{2}(q_{2})}\{\frac{1}{2m}(p_{1}^{2} + 
p_{2}^{2}) + U_{1}(q_{1}) + U_{2}(q_{2})\}
\end{equation}
where the variables change over $\infty > q_{1}, q_{2} >- \infty$. A general Hamiltonian of Liouville-type is given by 
\begin{equation}
H= \frac{1}{V_{1}(Q_{1}) + V_{2}(Q_{2})}\{\frac{1}{2m W_{1}(Q_{1})}P_{1}^{2} + 
\frac{1}{2m W_{2}(Q_{2})}P_{2}^{2} + U_{1}(Q_{1}) + U_{2}(Q_{2})\}
\end{equation}
but after a canonical transformation 
\begin{eqnarray}
\frac{1}{\sqrt{W_{1}(Q_{1})}}P_{1}= p_{1} &,& \int_{0}^{Q_{1}}\sqrt{W_{1}(Q)}dQ 
= q_{1}\nonumber\\
\frac{1}{\sqrt{W_{2}(Q_{2})}}P_{2}= p_{2} &,& \int_{0}^{Q_{2}}\sqrt{W_{2}(Q)}dQ 
= q_{2} 
\end{eqnarray}
and a suitable redefinition of $V$ and $U$, we can derive a Hamiltonian of the form in (2.1).

We may then solve the Schroedinger problem
\begin{equation}
E\psi = \frac{1}{V_{1}(q_{1}) + V_{2}(q_{2})}\{\frac{1}{2m}(\hat{p}_{1}^{2} + 
\hat{p}_{2}^{2}) + U_{1}(q_{1}) + U_{2}(q_{2})\}\psi 
\end{equation}
where
\begin{equation}
\hat{p}_{l} = -i\hbar\frac{\partial}{\partial q_{l}}
\end{equation}
for $l=1, 2$,and the volume element $dV$, which renders the Hamiltonian $H$
in (2.4) hermitian, is given by
\begin{equation}
dV = (V_{1}(q_{1}) + V_{2}(q_{2}))dq_{1}dq_{2}
\end{equation}
The classical Hamiltonian (2.1) does not completely specify the operator ordering in (2.4), and the simplest ordering is adopted here. A precise operator ordering needs to be fixed depending on each 
explicit example; a concrete example shall be given for the hydrogen atom later.

One may rewrite the above Schroedinger equation (2.4) as 
\begin{equation}
\hat{H}_{T} \psi = 0
\end{equation}
with a {\em total Hamiltonian} defined by a specific gauge condition, 
\begin{equation}
\hat{H}_{T} = \frac{1}{2m}(\hat{p}_{1}^{2} + \hat{p}_{2}^{2}) + U_{1}(q_{1}) + U_{2}(q_{2}) - E(V_{1}(q_{1}) + V_{2}(q_{2}))
\end{equation}
The meaning of a total Hamiltonian is clarified later.
A general procedure to deal with a completely separated operator $\hat{H}_{T}$ is to 
consider an evolution operator for a parameter $\tau$ defined by
\begin{eqnarray}
\langle q_{1b},q_{2b}| e^{- i\hat{H}_{T}\tau/\hbar} | q_{1a}, q_{2a}\rangle
&=& \langle q_{1b}| \exp{[- (i/\hbar)(\frac{1}{2m}\hat{p}_{1}^{2}  + U_{1}(q_{1}) 
          - EV_{1}(q_{1}))\tau]} | q_{1a}\rangle\nonumber\\
&\times&\langle q_{2b}| \exp{[- (i/\hbar)(\frac{1}{2m}\hat{p}_{2}^{2}  + U_{2}(q_{2}) 
          - EV_{2}(q_{2}))\tau]} | q_{2a}\rangle\nonumber\\
&=& \int{\cal D}q_{1}{\cal D}p_{1}e^{(i/\hbar)\int_{0}^{\tau}\{ p_{1}\dot{q_{1}} 
- (\frac{1}{2m}p_{1}^{2}  + U_{1}(q_{1}) - EV_{1}(q_{1}))\}d\tau}\nonumber\\
&\times&\int{\cal D}q_{2}{\cal D}p_{2}e^{(i/\hbar)\int_{0}^{\tau}\{ p_{2}\dot{q_{2}} - (\frac{1}{2m}p_{2}^{2}  + U_{2}(q_{2}) - EV_{2}(q_{2}))\}d\tau}
\end{eqnarray}
The parameter $\tau$ is arbitrary , and by integrating over $\tau$ from $0$ to $\infty$ one obtains a physically meaningful quantity
\begin{eqnarray}
&&\langle q_{1b},q_{2b}|\frac{\hbar}{\hat{H}_{T}} | q_{1a}, q_{2a}\rangle_{semi-classical}
\nonumber\\
&=&i \int_{0}^{\infty}d\tau \frac{1}{\sqrt{2\pi i\hbar (\partial q_{1}(\tau)/
\partial p_{1}(0))_{q_{1a}}}}\frac{1}{\sqrt{2\pi i\hbar (\partial q_{2}(\tau)/
\partial p_{2}(0))_{q_{2a}}}}\nonumber\\
&\times&\exp{\{(i/\hbar)S_{cl}(q_{1b},q_{1a},\tau) + (i/\hbar)S_{cl}(q_{2b},q_{2a},\tau)\}}
\end{eqnarray} 
where we wrote the result of a semi-classical approximation for the path 
integral[17][18] [19], though in certain cases one may be able to perform an exact path integral in (2.9). The pre-factor  in (2.10) is written in terms of  classical 
paths , for example,
\begin{equation}
q_{1cl}(\tau) = q_{1}(\tau; q_{1a}, p_{1}(0))
\end{equation}
Namely, the classical paths dictated by the total Hamiltonian $\hat{H}_{T}$ are
expressed as functions of the initial positions and momenta. On the other hand,
the classical
action $S_{cl}$ is expressed as a function of the initial position, final
position and elapsed ``time'' $\tau$ by eliminating $p_{1}(0)$ dependence;for 
example, 
\begin{equation}
S_{cl}(q_{1b},q_{1a},\tau) = \int_{0}^{\tau}\{ p_{1}\dot{q_{1}} 
- (\frac{1}{2m}p_{1}^{2}  + U_{1}(q_{1}) - EV_{1}(q_{1}))\}_{cl}d\tau
\end{equation}
with $q_{1}(\tau) = q_{1b}$. If one solves the Hamilton-Jacobi equation in the 
form
\begin{equation}
S(q_{1b}, q_{1a}; \tau) = - A\tau + S(q_{1b}, q_{1a}; A )
\end{equation}
one treats $A$ as a dynamical variable and regards the above equation as a Legendre transformation defined by
\begin{eqnarray}
\frac{\partial S(q_{1b}, q_{1a}; \tau)}{\partial \tau} &=& -A\nonumber\\
\frac{\partial S(q_{1b}, q_{1a}; A )}{\partial A} & = & \tau
\end{eqnarray}
The variable $A$ is then eliminated.  This may be regarded as a classical analogue of uncertainty
relation; if one specifies $\tau$, the conjugate variable $A$ becomes implicit. It is known that the semi-classical approximation (2.10) is exact for a 
quadratic system such as a simple harmonic oscillator[18].

We next note the relation for the quantity defined in the left-hand side of (2.9) 
\begin{eqnarray}
&&\langle q_{1b},q_{2b}|\frac{1}{\hat{H}_{T}} | q_{1a}, q_{2a}\rangle
\nonumber\\
&=&\langle q_{1b},q_{2b}|\frac{1}{(\frac{1}{\hat{V}_{1}(q_{1}) +
\hat{V}_{2}(q_{2})})\hat{H}_{T}} | q_{1a}, q_{2a}\rangle\frac{1}{V_{1}(q_{1a}) + V_{2}(q_{2a})}\nonumber\\
&=&\langle q_{1b},q_{2b}|\frac{1}{\hat{H} - E} | q_{1a}, q_{2a}\rangle\frac{1}{V_{1}(q_{1a}) + V_{2}(q_{2a})}\\
&=&\frac{1}{H( q_{1b},\frac{\hbar}{i}\frac{\partial}{\partial q_{1b}}, ..) - E}\{\frac{1}{\sqrt{V_{1}(q_{1b}) + V_{2}(q_{2b})}}\langle q_{1b},q_{2b}| q_{1a}, q_{2a}\rangle\frac{1}{\sqrt{V_{1}(q_{1a}) + V_{2}(q_{2a})}}\}\nonumber
\end{eqnarray}
by recalling $(\hat{A}\hat{B})^{-1} =\hat{B}^{-1}\hat{A}^{-1}$. The state vectors in these 
relations are defined for the volume element $dq_{1}dq_{2}$ as
\begin{equation}
\int dq_{1}dq_{2} |q_{1},q_{2}\rangle \langle q_{1},q_{2}| = 1
\end{equation}
Note that the definition of the $\delta$-function in $\langle q_{1}^{\prime},q_{2}^{\prime}|q_{1},q_{2}\rangle = \delta (q_{1}^{\prime}- q_{1})\delta (q_{2}^{\prime}- q_{2})$ depends on the choice of the volume elememt in (2.16) and thus on the choice of $H_{T}$.
The last expression in (2.15) is thus correctly defined for the original
Hamiltonian $H$ and the original state $\psi$ in (2.4)  with the volume element $dV$
in (2.6), since we have the completeness relation from (2.16)
\begin{equation}
\int  |q_{1},q_{2}\rangle \frac{dV}{V_{1}(q_{1}) + V_{2}(q_{2})}\langle q_{1},q_{2}| = 1 
\end{equation}
The left-hand side of (2.15) thus defines the correct Green's function 
for the original operator $(\hat{H} - E)^{-1}$ by noting the symmetry in $q_{a}$ and $q_{b}$.  One can then define the conventional evolution operator by 
\begin{eqnarray}
&&\langle q_{1b},q_{2b}|e^{ -i\hat{H}(t_{b} - t_{a})/\hbar} | q_{1a}, q_{2a}\rangle _{conv}\nonumber\\
&& = \frac{1}{2\pi i\hbar}\int_{-\infty}^{\infty} dE e^{-i E(t_{b} - t_{a})/\hbar}\langle q_{1b},q_{2b}|\frac{\hbar}{\hat{H}-i\epsilon - E} | q_{1a}, q_{2a}\rangle\frac{1}{V_{1}(q_{1a}) + V_{2}(q_{2a})}
\end{eqnarray}
where $\epsilon$ is an infinitesimal positive number. 
The total Hamiltonian changes for a different choice of gauge condition in the 
Jacobi's principle of least action to be explained below. Consequently, the 
volume element, which renders $H_{T}$ hermitian, generally depends on the 
choice of gauge. In this case, one has the relation 
\begin{eqnarray}
\langle q_{1b},q_{2b} | q_{1a}, q_{2a}\rangle_{conv} &=&\langle q_{1b},q_{2b} | q_{1a}, q_{2a}\rangle\frac{1}{V_{1}(q_{1a}) + V_{2}(q_{2a})}\nonumber\\
 &=& \frac{1}{\sqrt{V_{1}(q_{1b}) + V_{2}(q_{2b})}}\langle q_{1b},q_{2b}|q_{1a}, q_{2a}\rangle\frac{1}{\sqrt{V_{1}(q_{1a}) + V_{2}(q_{2a})}} \nonumber
\end{eqnarray}

The meaning of the total Hamiltonian $H_{T}$ (2.8) becomes transparent if one 
starts with the Jacobi's principle of least action for a given $E$
\begin{eqnarray}
S& =& \int_{0}^{\tau} d\tau L = \int_{0}^{\tau}d\tau \sqrt{2m[E(V_{1}(q_{1}) + V_{2}(q_{2})) - (U_{1}(q_{1}) + U_{2}(q_{2}))](\dot{q}_{1}^{2} + \dot{q}_{2}^{2})}\nonumber\\
&=& \int \sqrt{2m[E(V_{1}(q_{1}) + V_{2}(q_{2})) - (U_{1}(q_{1}) + U_{2}(q_{2}))][(dq_{1})^{2}+ (dq_{2})^{2}]}
\end{eqnarray}
which is reparametrization invariant. One then defines the momenta conjugate to coordinates
\begin{equation}
p_{l} = \frac{\partial L}{\partial \dot{q}_{l}} = \sqrt{2m[E(V_{1}(q_{1}) + V_{2}(q_{2})) - (U_{1}(q_{1}) + U_{2}(q_{2}))]}\times 
\frac{\dot{q}_{l}}{\sqrt{ (\dot{q}_{1}^{2} + \dot{q}_{2}^{2})}}
\end{equation}
and obtains a vanishing Hamiltonian, which is a result of reparametrization 
invariance,   and a first class constraint $\phi$ as the generator of reparametrization gauge symmetry,
\begin{eqnarray}
H &=& p_{l}\dot{q}_{l} - L = 0\nonumber\\
\phi(q_{l}, p_{l}) &=& \frac{1}{V_{1}(q_{1}) + V_{2}(q_{2})}\{\frac{1}{2m}
(p_{1}^{2} + p_{2}^{2}) + U_{1}(q_{1}) + U_{2}(q_{2})\} - E \simeq 0
\end{eqnarray}

Following Dirac[20], one may then define a total Hamiltonian
\begin{eqnarray}
H_{T}& =& H + \alpha (q_{l}, p_{l}) \phi (q_{l}, p_{l})\nonumber\\
     &=& \alpha (q_{l}, p_{l}) \phi (q_{l}, p_{l}) \simeq 0
\end{eqnarray}
where an arbitrary function $\alpha (q_{l}, p_{l})$ specifies a choice of gauge or a choice of the arbitrary parameter $\tau$ in (2.19), which parametrizes the orbit  for a given $E$.  The quantum theory is defined by( up to an operator ordering)
\begin{equation}
 i\hbar \frac{\partial}{\partial \tau}\psi =  \hat{H}_{T}\psi
\end{equation}
with a physical state condition
\begin{equation}
 \hat{ \alpha} (q_{l}, p_{l})\hat{ \phi} (q_{l}, p_{l})\psi_{phy} = 0
\end{equation}
A choice of the specific gauge  $\alpha (q_{l}, p_{l}) = V_{1}(q_{1}) + V_{2}(q_{2})$ gives rise to (2.7) and the choice $\alpha (q_{l}, p_{l}) = 1 $ gives the conventional static Schroedinger equation (2.4), since $\psi$ appearing in these 
equations are physical states.

The basic dynamical principle involved is thus identified as the Jacobi's principle of least 
action, which is analogous to geometrical optics, and the formula  of an evolution operator (2.9) dictated by (2.23) provides a basis for the path integral approach 
to a general separable Hamiltonian of Liouville-type. The path integral in (2.9) deals with a sum over orbits in space instead of space-time, and the notion of re-scaled time does not explicitly appear in the present approach; the evolution operator (2.9) essentially generates a gauge transformation.

\section{Hydrogen Atom}

{\bf 3.1, Analysis in Parabolic Coordinates}

We analyze the hydrogen atom by starting  with the Hamiltonian written in terms of   parabolic coordinates
\begin{equation}
H(\xi, \eta, \varphi)  = \frac{1}{2m(\xi + \eta)}(\xi p_{\xi}^{2} + \eta p_{\eta}^{2}) + \frac{1}
{8m\xi \eta}p_{\varphi}^{2} - \frac{e^{2}}{\xi + \eta}
\end{equation}
where the parabolic coordinates are introduced via the cylindrical
coordinates $(\rho, \varphi, z)$    by
\begin{eqnarray}
\xi = \frac{1}{2}(r - z)\nonumber\\
\eta =\frac{1}{2}(r + z)\nonumber\\
r = \sqrt{\rho^{2} + z^{2}}
\end{eqnarray}
and $\varphi$ stands for the azimuthal angle around the $z$ axis. We further 
perform a canonical transformation which simplifies the kinetic term in $H$
as
\begin{eqnarray}
\xi &=& \frac{1}{4}u^{2}\ \ \ \ \ \ \ \ \ ,0\leq u < \infty \nonumber\\
\sqrt{\xi}p_{\xi}&=& p_{u}\nonumber\\
\eta &=& \frac{1}{4}v^{2}\ \ \ \ \ \ \ \ \ ,0\leq v < \infty \nonumber\\
\sqrt{\eta}p_{\eta}&=& p_{v}
\end{eqnarray}
and the Hamiltonian becomes
\begin{equation}
H = \frac{1}{2m}(\frac{4}{u^{2} + v^{2}})[p_{u}^{2} + \frac{1}{u^{2}}p_{\varphi}^{2} + p_{v}^{2} + \frac{1}{v^{2}}p_{\varphi}^{2}] 
- \frac{4}{u^{2} + v^{2}}e^{2}
\end{equation}
where $ r = \xi + \eta = (u^{2} + v^{2})/4$. This Hamiltonian is not yet a separable one of Liouville-type.

One may solve the Schroedinger equation 
\begin{equation}
\hat{H} \psi = E\psi
\end{equation}
or equivalently
\begin{equation}
\hat{H}_{T}\psi = 0
\end{equation}
with
\begin{equation}
\hat{H}_{T} = \frac{1}{2m}[{\hat{p}_{u}}^{2} + \frac{1}{u^{2}}{\hat{p}_{\varphi}}^{2} + {\hat{p}_{v}}^{2}+  \frac{1}{v^{2}}{\hat{p}_{\varphi}}^{2}] -e^{2} 
+ \frac{m\omega^{2}}{2}(u^{2} +v^{2})
\end{equation}
where $\omega$ is defined by
\begin{equation}
\frac{1}{2}m{\omega}^{2} = -\frac{1}{4}E
\end{equation}
We consider the case $E < 0$ for the moment.
$\hat{H}_{T}$ stands for the total Hamiltonian defined by a specific gauge 
condition; a  general definition of $\hat{H}_{T}$ will
be given later in (3.47).  

Eq.(3.6) may be rewritten in an equivalent form as
\begin{eqnarray}
\hat{\tilde{H}}_{T}\psi &=& 0\nonumber\\
(\hat{p}_{\varphi} -  \hat{p}_{{\varphi}^{\prime}})\psi& =& 0
\end{eqnarray}
We here introduced auxiliary variables  $(\hat{p}_{{\varphi}^{\prime}},
\varphi^{\prime})$ as
\begin{eqnarray}
\hat{\tilde{H}}_{T}&=& \frac{1}{2m}[\hat{p}_{u}^{2} + \frac{1}{u^{2}}\hat{p}_{\varphi}^{2} + \hat{p}_{v}^{2} +  \frac{1}{v^{2}}\hat{p}_{{\varphi}^{\prime}}]  + \frac{m\omega^{2}}{2}(u^{2} +v^{2}) - e^{2}\nonumber\\
&=&\frac{1}{2m}\vec{p}_{u}^{2} + \frac{m\omega^{2}}{2}\vec{u}^{2} +
   \frac{1}{2m}\vec{p}_{v}^{2} + \frac{m\omega^{2}}{2}\vec{v}^{2} - e^{2}
\end{eqnarray}
and we  defined 
\begin{eqnarray}
\vec{u} &=& ( u_{1},u_{2})= (u\cos\varphi, u\sin\varphi)\nonumber\\
\vec{p}_{u}^{2} &=& \hat{p}_{u}^{2} + \frac{1}{u^{2}}\hat{p}_{\varphi}^{2}\nonumber\\
\vec{v} &=& ( v_{1},v_{2})= (v\cos{\varphi^{\prime}},v\sin{\varphi^{\prime}})\nonumber\\
\vec{p}_{v}^{2} &=& \hat{p}_{v}^{2} + \frac{1}{v^{2}}\hat{p}_{{\varphi}^{\prime}}^{2}
\end{eqnarray} 
The subsidiary condition in (3.9) replaces the use of the Kustaanheimo-Stiefel transformation, and at the same time it renders a Hamiltonian of Liouville-type. This introduction of auxiliary variables (3.11) has been discussed by Ravndal and Toyoda[16]. The use of the subsidiary condition (3.9) in place of the
Kustaanheimo-Stiefel transformation may be useful for a pedagogical purpose. 

A general procedure to deal with a completely separated operator such as $\hat{\tilde{H}}_{T}$ in (3.10) is to consider an evolution operator for a parameter
$\tau$ defined by
\begin{eqnarray}
&&\langle \vec{u}_{b}, \vec{v}_{b}|e^{-i\hat{\tilde{H}}_{T}\tau/\hbar}|\vec{u}_{a}, \vec{v}_{a}\rangle\nonumber\\
&&= e^{ie^{2}\tau}\langle \vec{u}_{b}|exp[ -(i/\hbar)(\frac{1}{2m}\vec{p}_{u}^{2} + \frac{m\omega^{2}}{2}\vec{u}^{2})\tau]|\vec{u}_{a}\rangle\nonumber\\
&&\times \langle \vec{v}_{b}|exp[ -(i/\hbar)(\frac{1}{2m}\vec{p}_{v}^{2} + \frac{m\omega^{2}}{2}\vec{v}^{2})\tau]|\vec{v}_{a}\rangle\nonumber\\
&&= e^{ie^{2}\tau}(\frac{m\omega}{2\pi i\hbar \sin \omega \tau})^{4/2}\nonumber\\
&&\times exp\{\frac{im\omega}{2\hbar \sin \omega\tau}[(\vec{u}_{b}^{2} +\vec{v}_{b}^{2} + \vec{u}_{a}^{2} + \vec{v}_{a}^{2})\cos \omega\tau -2\vec{u}_{b}\vec{u}_{a} -
2\vec{v}_{b}\vec{v}_{a}]\}
\end{eqnarray}
where we used the exact result for a simple harmonic oscillator[18]
\begin{eqnarray}
&&\langle q_{b}|exp[ -(i/\hbar)(\frac{1}{2m}\hat{p}^{2} + \frac{m\omega^{2}}{2}\hat{q}^{2})\tau]|q_{a}\rangle\nonumber\\
&& = (\frac{m\omega}{2\pi i\hbar\sin\omega\tau})^{1/2}
exp\{ \frac{im\omega}{2\hbar\sin\omega\tau}[(q_{b}^{2} + q_{a}^{2})\cos\omega\tau -2 q_{b}q_{a}]\}
\end{eqnarray}
which can be established either in path integral or in operator formalism( see
(2.10)). Also, the harmonic oscillator is symmetric with respect to coordinate 
or momentum space  representation [2].

A crucial observation here is that $\hat{p}_{\varphi}$ and $\hat{p}_{{\varphi}^{\prime}}$ are preserved during the evolution dictated by the operator $\hat{\tilde{H}}_{T}$ in (3.10), since $ [\hat{p}_{\varphi}, \hat{\tilde{H}}_{T}] =  
[\hat{p}_{{\varphi}^{\prime}}, \hat{\tilde{H}}_{T}] =  0$. It is then sufficient to impose the constraint (3.9) only on the initial state , for example. 
Starting with a general state belonging to the eigenvalues $\hat{p}_{\varphi}=
m$ and $\hat{p}_{{\varphi}^{\prime}} = m^{\prime}$ 
\begin{equation}
e^{i m\varphi} e^{i m^{\prime}\varphi^{\prime}} 
\end{equation}
we can use the following trick 
\begin{equation}
\int_{0}^{2\pi} \frac{d\theta}{2\pi} e^{i m(\varphi + \theta)}e^{i m^{\prime}
(\varphi^{\prime} - \theta)} = \delta_{m,m^{\prime}}e^{i m(\varphi + \varphi^{\prime})}
\end{equation}
to project out the state satisfying $\hat{p}_{\varphi}= \hat{p}_{{\varphi}^{\prime}}$, and $\varphi + \varphi^{\prime}$ is regarded as the actual azimuthal
angle.

We thus obtain
\begin{eqnarray}
&&\langle u_{b}, v_{b}, (\varphi + \varphi^{\prime})_{b}|e^{-i \hat{H}_{T}\tau/\hbar} | u_{a}, v_{a}, (\varphi + \varphi^{\prime})_{a}\rangle\nonumber\\
&&= e^{ie^{2}\tau}(\frac{m\omega}{2\pi i\hbar \sin \omega \tau})^{2}
\int_{0}^{2\pi}\frac{d\theta}{2\pi}exp\{\frac{im\omega}{2\hbar \sin \omega\tau}[(\vec{u}_{b}^{2} +\vec{v}_{b}^{2} + \vec{u}_{a}^{2} + \vec{v}_{a}^{2})\cos \omega\tau -2\vec{u}_{b}\vec{u}_{a} -
2\vec{v}_{b}\vec{v}_{a}]\}\nonumber\\
&&= e^{ie^{2}\tau}(\frac{m\omega}{2\pi i\hbar \sin \omega \tau})^{2}\times \nonumber\\
&&\int_{0}^{2\pi}\frac{d\theta}{2\pi}exp\{\frac{im\omega}{2\hbar \sin \omega\tau}[4(\xi_{a} + \xi_{b} + \eta_{a} +
\eta_{b} )\cos \omega\tau - 4\sqrt{2}(r_{a}r_{b} + \vec{x}_{a}\vec{x}_{b})^{1/2}\cos (\theta + \gamma) ]\}\nonumber\\
&&=  e^{ie^{2}\tau}(\frac{m\omega}{2\pi i\hbar \sin \omega \tau})^{2}
exp\{\frac{2 im\omega}{\hbar \sin \omega\tau}
(r_{a} + r_{b})\cos \omega\tau\}
I_{0}(\frac{2\sqrt{2} i m\omega}{\hbar \sin \omega \tau}(r_{a}r_{b} + \vec{x}_{a}\vec{x}_{b})^{1/2})\nonumber\\
&& 
\end{eqnarray}
In this evaluation we start with the relation
\begin{equation}
\vec{u}_{b}\vec{u}_{a} + \vec{v}_{b}\vec{v}_{a}
= u_{b}u_{a}\cos \Delta\varphi + v_{b}v_{a}\cos \Delta\varphi^{\prime}
\end{equation}
with $\Delta\varphi = \varphi_{b} - \varphi_{a}, \Delta\varphi^{\prime} =
\varphi^{\prime}_{b} - \varphi^{\prime}_{a}$, and  
\begin{eqnarray}
&&u_{b}u_{a}\cos (\Delta\varphi+\theta) + v_{b}v_{a}\cos (\Delta\varphi^{\prime}-\theta)\nonumber\\
&&= (u_{b}u_{a}\cos \Delta\varphi + v_{b}v_{a}\cos \Delta\varphi^{\prime})
\cos\theta \nonumber\\
&&+(- u_{b}u_{a}\sin \Delta\varphi + v_{b}v_{a}\sin \Delta\varphi^{\prime})
\sin\theta \nonumber\\ 
&&=4\sqrt{\xi_{b}\xi_{a} +\eta_{b}\eta_{a} +2\sqrt{\xi_{b}\xi_{a}\eta_{b}\eta_{a}}\cos(\Delta\varphi + \Delta\varphi^{\prime})}\cos(\theta +\gamma)\nonumber\\
&&=2\sqrt{2}\sqrt{r_{a}r_{b} + z_{a}z_{b} + \rho_{a}\rho_{b}\cos(\Delta\varphi + \Delta\varphi^{\prime})}\cos(\theta +\gamma)\nonumber\\
&&=2\sqrt{2}\sqrt{r_{a}r_{b} + \vec{x}_{a}\vec{x}_{b}}\cos(\theta +\gamma)
\end{eqnarray}
where we used the definition of  variables in (3.3),  and $\gamma$ is a number independent of $\theta$. We also defined a modified Bessel function
\begin{equation}
I_{0}(\frac{2\sqrt{2} i m\omega}{\hbar \sin \omega \tau}(r_{a}r_{b} + \vec{x}_{a}\vec{x}_{b})^{1/2}) = \int_{0}^{2\pi}\frac{d\theta}{2\pi}exp\{\frac{2\sqrt{2}im\omega (r_{a}r_{b} + \vec{x}_{a}\vec{x}_{b})^{1/2}   }{\hbar \sin \omega\tau}\cos \theta  \}
\end{equation}

The parameter $\tau$ is arbitrary, and we eliminate $\tau$ to obtain a physically meaningful quantity by 
\begin{eqnarray}
&&i\int_{0}^{\infty} d\tau \langle u_{b}, v_{b}, (\varphi + \varphi^{\prime})_{b}|e^{-i \hat{H}_{T}\tau/\hbar} | u_{a}, v_{a}, (\varphi + \varphi^{\prime})_{a}\rangle\nonumber\\
&&=\langle u_{b}, v_{b}, (\varphi + \varphi^{\prime})_{b}|\frac{\hbar}{\hat{H}_{T}} | u_{a}, v_{a}, (\varphi + \varphi^{\prime})_{a}\rangle\nonumber\\
&&= i\int_{0}^{\infty} d\tau 
e^{ie^{2}\tau}(\frac{m\omega}{2\pi i\hbar \sin \omega \tau})^{2}
exp\{\frac{2 im\omega}{\hbar \sin \omega\tau}
(r_{a} + r_{b})\cos \omega\tau\}\nonumber\\
&&\times I_{0}(\frac{2\sqrt{2} i m\omega}{\hbar \sin \omega \tau}(r_{a}r_{b} + \vec{x}_{a}\vec{x}_{b})^{1/2})\nonumber\\
&&=\frac{m\omega}{2\pi^{2}\hbar^{2}}\int_{0}^{1}d\lambda \lambda^{-\nu}
\frac{1}{(1-\lambda)^{2}}
exp [ \frac{-2m\omega}{\hbar}(r_{a} + r_{b})(\frac{1+\lambda}{1 - \lambda})]I_{0}(\frac{4\sqrt{2}m\omega}{\hbar}
\frac{\lambda^{1/2}
}{1-\lambda}(r_{a}r_{b} + \vec{x}_{a}\vec{x}_{b})^{1/2})
\nonumber\\
&&
\end{eqnarray}
where we rotated $\tau$ by 90 degrees , $\tau \rightarrow -i\tau$, and defined
\begin{eqnarray}
\lambda &=& e^{-2\omega \tau}\nonumber\\
\nu &=& e^{2}/2\omega
\end{eqnarray}

We next show that (3.20) gives an exact Green's function for the hydrogen atom by noting the sequence
\begin{eqnarray}
&&\langle u_{b}, v_{b}, \varphi_{b}|\frac{\hbar}{\hat{H}_{T}} | u_{a}, v_{a}, 
\varphi_{a}\rangle\nonumber\\
&&=\langle \xi_{b}, \eta_{b}, \varphi_{b}|\frac{\hbar}{(\frac{1}{\hat{\xi} + \hat{\eta}})\hat{H}_{T}(\xi, \eta, \varphi)} | \xi_{a}, \eta_{a}, 
\varphi_{a}\rangle (\frac{1}{\xi_{a} + \eta_{a}})\nonumber\\
&&=\langle \xi_{b}, \eta_{b}, \varphi_{b}|\frac{\hbar}{\hat{H}(\xi, \eta, \varphi) - E} | \xi_{a}, \eta_{a}, 
\varphi_{a}\rangle (\frac{1}{\xi_{a} + \eta_{a}})\nonumber\\
&&=\frac{\hbar}{\hat{H}(\xi_{b},\eta_{b},\varphi_{b}) - E}(\frac{1}{\sqrt{\xi_{b}+ \eta_{b}}}\langle \xi_{b}, \eta_{b}, \varphi_{b}| \xi_{a}, \eta_{a}, 
\varphi_{a}\rangle \frac{1}{\sqrt{\xi_{a} + \eta_{a}}})\nonumber\\
&&= \frac{1}{4\pi}\langle \vec{x}_{b}|\frac{\hbar}{\hat{\vec{p}}^{2}/{2m}
- e^{2}/r - E}|\vec{x}_{a}\rangle
\end{eqnarray}
where we used $\varphi$ in place of $\varphi + \varphi^{\prime}$ and the relation $(\hat{A}\hat{B})^{-1} = \hat{B}^{-1}\hat{A}^{-1}$.

The volume element changes in this transition from $\hat{H}_{T}$ to $\hat{H}$ as
\begin{eqnarray}
dV_{0} &=& 2\pi uvdudvd\varphi\nonumber\\
\rightarrow  dV &=& (\xi + \eta)dV_{0} = 4\pi\times 2(\xi + \eta)d\xi d\eta d\varphi\nonumber\\
   &=& 4\pi\times  r^{2}dr d\cos \theta d\varphi
\end{eqnarray}
The bra- and ket- vectors in (3.22) are normalized in the combination 
\begin{eqnarray}
\int dV_{0}| u, v, \varphi\rangle \langle u, v, \varphi| &=& 1\nonumber\\
\int dV| \xi, \eta, \varphi\rangle \frac{1}{\xi + \eta} \langle \xi, \eta, \varphi| &=& 1\nonumber\\
\int d^{3}x |\vec{x}\rangle \langle \vec{x}| &=& 1
\end{eqnarray}
and the extra factor of $4\pi$ in $dV = 4\pi r^{2}dr d\cos \theta d\varphi$ requires
the appearance of the factor of $1/4\pi$ in the last expression  in (3.22). The appearance of $2\pi$ in $dV_{0}$ is an artifact of the variable $\varphi^{\prime}$ in (3.11). This normalization condition of bra-  and ket- vectors together 
with a symmetry in $\vec{x}_{a}$ and $\vec{x}_{b}$ justify the identification
(3.22). A more explicit and concrete analysis of eqs.(3.22)$\sim $
 (3.24) will be given in connection with the Jacobi's principle later.

As for the operator ordering, the momentum operator  changes in (3.22) as
\begin{eqnarray}
\hat{p}^{2}_{u} + \hat{p}^{2}_{v} &=& (\frac{\hbar}{i})^{2}[\frac{1}{u}\partial
_{u}u\partial_{u} + \frac{1}{v}\partial_{v}v\partial_{v}]\nonumber\\
&=&(\frac{\hbar}{i})^{2}[\partial_{\xi}\xi\partial_{\xi} + \partial_{\eta}\eta\partial_{\eta}]\nonumber\\ 
&=& \hat{p}_{\xi}\xi\hat{p}_{\xi} + \hat{p}_{\eta}\eta\hat{p}_{\eta}
\end{eqnarray}
and 
\begin{equation}
(\frac{1}{\xi + \eta})( \hat{p}_{\xi}\xi \hat{p}_{\xi} + \hat{p}_{\eta}\eta \hat{p}_{\eta} ) + \frac{1}{4\xi\eta}{\hat{p}_{\varphi}}^{2} ={ \hat{\vec{p}}}^{2}
\end{equation}
where the right-hand side is written in cartesian coordinates.
We note that $dV_{0}$ and $dV$ in (3.23) respectively render $\hat{H}_{T}$ and 
$\hat{H}(\xi, \eta, \varphi )$ hermitian.

Combining (3.20),(3.22) and (3.26), we have thus established the exact Green's 
function including the operator ordering 
\begin{eqnarray}
\langle \vec{x}_{b}|\frac{\hbar}{\hat{\vec{p}}^{2}/{2m}
- e^{2}/r - E}|\vec{x}_{a}\rangle &=& 
\frac{2 m^{2}\omega}{\pi \hbar^{2}}\int_{0}^{1}d\lambda \lambda^{-\nu}
\frac{1}{(1-\lambda)^{2}}
exp [ \frac{-2m\omega}{\hbar}(r_{a} + r_{b})(\frac{1+\lambda}{1 - \lambda})]
\nonumber\\
&& \times I_{0}(\frac{4\sqrt{2}m\omega}{\hbar}
\frac{\lambda^{1/2}}{1-\lambda}(r_{a}r_{b} + \vec{x}_{a}\vec{x}_{b})^{1/2})
\end{eqnarray}
It is known that this formula, which was first derived by Duru and 
Kleinert[4], is a Fourier transform of Schwinger's momentum space representation[1]. The continuation to the scattering problem with $E > 0$ is performed by
the replacement
\begin{equation}
\omega \rightarrow (-i)\omega, \ \ \nu \rightarrow i\nu
\end{equation}
in the above formula.

One can understand the spectrum of the hydrogen atom by looking at $\tilde{H}_{T}$ in (3.10). This problem, which has been analyzed by Ravndal and Toyoda[16], is briefly summarized here in connection with the Jacobi's
principle of least action and the Stark effect to be discussed later.
If one defines the oscillator variables 
\begin{eqnarray}
a_{k} &=& \frac{1}{\sqrt{2}}[\sqrt{\frac{m\omega}{\hbar}} u_{k} + 
          \frac{i}{\sqrt{m\omega \hbar}}\hat{p}_{u_{k}}], \nonumber\\
\tilde{a}_{k} &=& \frac{1}{\sqrt{2}}[\sqrt{\frac{m\omega}{\hbar}} v_{k} + 
          \frac{i}{\sqrt{m\omega \hbar}}\hat{p}_{v_{k}}],\ \ k = 1, 2
\end{eqnarray}
one obtains 
\begin{eqnarray}
\hat{\tilde{H}}_{T} &=& \hbar\omega [\sum_{k=1}^{2}( a_{k}^{\dagger}a_{k} 
            + \tilde{a}_{k}^{\dagger}\tilde{a}_{k}) + 2 ] - e^{2}\nonumber\\
\hat{p}_{\varphi} &=& i\hbar [a_{1}^{\dagger}a_{2} - a_{2}^{\dagger}a_{1}]
\nonumber\\
\hat{p}_{\varphi^{\prime}} &=& i\hbar [\tilde{a}_{1}^{\dagger}\tilde{a}_{2} - 
\tilde{a}_{2}^{\dagger}\tilde{a}_{1}]
\end{eqnarray}
After a unitary transformation
\begin{eqnarray}
a_{1} &=& \frac{1}{\sqrt{2}}( b_{1} - i b_{2})\nonumber\\
a_{2} &=& \frac{1}{\sqrt{2}}( - i b_{1} +  b_{2})
\end{eqnarray}
and a similar transformation of $\tilde{a}_{1}$ and $\tilde{a}_{2}$, one 
obtains
\begin{eqnarray}
\hat{\tilde{H}}_{T} &=& \hbar\omega [\sum_{k=1}^{2}( b_{k}^{\dagger}b_{k} 
            + \tilde{b}_{k}^{\dagger}\tilde{b}_{k}) + 2 ] - e^{2}\nonumber\\
\hat{p}_{\varphi} &=& \hbar [b_{1}^{\dagger}b_{1} - b_{2}^{\dagger}b_{2}]
\nonumber\\
\hat{p}_{\varphi^{\prime}} &=& \hbar [\tilde{b}_{1}^{\dagger}\tilde{b}_{1} - 
\tilde{b}_{2}^{\dagger}\tilde{b}_{2}]
\end{eqnarray}
By defining the number operators
\begin{eqnarray}
n_{k} &=& b^{\dagger}_{k}b_{k},\nonumber\\
\tilde{n}_{k} &=& \tilde{b}^{\dagger}_{k}\tilde{b}_{k}, \ \ k = 1, 2
\end{eqnarray}
the total Hamiltonian is given by 
\begin{eqnarray}
\hat{\tilde{H}}_{T} &=& \hbar\omega [n_{1} + n_{2} + \tilde{n}_{1} + \tilde{n}_{2} + 2 ] - e^{2}\nonumber\\
&=& \hbar\omega [ 2n_{1} - \hat{p}_{\varphi}/\hbar + 2\tilde{n}_{2} + \hat{p}_{\varphi^{\prime}}/\hbar + 2 ] - e^{2}\nonumber\\
&=& 2\hbar\omega [ n_{1} + \tilde{n}_{2} + 1 ] - e^{2} 
\end{eqnarray}
by noting the physical state condition $\hat{p}_{\varphi}= \hat{p}_{\varphi^{\prime}}$.

We thus define the principal quantum number (or its operator) $n$ by 
\begin{equation}
n = n_{1} + \tilde{n}_{2} + 1 = 1, 2, 3, .....
\end{equation}
and the physical state condition 
\begin{equation}
( 2 n \hbar\omega - e^{2})\psi_{phys} = 0
\end{equation}
gives rise to the Bohr spectrum
\begin{equation}
E = - \frac{mc^{2}}{2} (\frac{e^{2}}{\hbar c})^{2}\frac{1}{n^{2}}, \ \ n= 1, 2, 3,...
\end{equation}
by noting the definition of $\omega$ in (3.8).

As for the degeneracy of  states with a fixed $n$, we have $n$ combinations of $(n_{1}, \tilde{n}_{2})$ given by 
\begin{eqnarray}
\tilde{n}_{2} &=& n - (n_{1} + 1 ),\nonumber\\
n_{1} &=& 0, 1, ......., n - 1
\end{eqnarray}
For each fixed $(n_{1}, \tilde{n}_{2})$, we have the constraints
\begin{eqnarray}
n_{2} &=& n_{1} - \hat{p}_{\varphi}/\hbar \geq  0\nonumber\\
\tilde{n}_{1} &=& \tilde{n}_{2} + \hat{p}_{\varphi^{\prime}}/\hbar = n - (n_{1} + 1) + \hat{p}_{\varphi^{\prime}}/\hbar \geq 0
\end{eqnarray}
which gives 
\begin{equation}
n_{1} \geq \hat{p}_{\varphi}/\hbar = \hat{p}_{\varphi^{\prime}}/\hbar \geq - n + (n_{1} + 1 )
\end{equation}
Namely, we have 
\begin{equation}
n_{1} - ( -n + n_{1} + 1  ) + 1 = n
\end{equation}
possible values of $\hat{p}_{\varphi}$. We thus have 
\begin{equation}
n\times n = n^{2}
\end{equation}
degeneracy of  states with a fixed  principal quantum number $n$, as required. This somewhat unorthodox classification of states may be useful in the analysis of the Stark effect 
to be commented on later.

Incidentally, the formula (3.32) suggests that one can define the Green's 
function by 
\begin{equation}
i\int_{0}^{\infty}d\tau\int_{0}^{2\pi}d\theta \langle f| exp\{ -\frac{i}{\hbar}\{ \hbar\omega [\sum_{k=1}^{2}( b_{k}^{\dagger}b_{k} 
            + \tilde{b}_{k}^{\dagger}\tilde{b}_{k}) + 2 ] - e^{2}\}\tau- i  (b_{1}^{\dagger}b_{1} - b_{2}^{\dagger}b_{2} 
            - \tilde{b}_{1}^{\dagger}\tilde{b}_{1} + \tilde{b}_{2}^{\dagger}\tilde{b}_{2})\theta \}|i\rangle
\end{equation}
for a direct path integral in (3.16), instead of using the trick noted in (3.15).  The $\theta$ integral imposes the constraint $ \hat{p}_{\varphi} = \hat{p}_{{\varphi}^{\prime}}$.  The path integral of an evolution operator in terms of oscillator variables is known as holomorphic ( or coherent state) path integral[21]. It is interesting that (3.43) resembles a closed string propagation on a 
cylindrical world-sheet.\\
{\bf 3.2, Jacobi's Principle of Least Action}

The meaning of the total Hamiltonian in (3.7) becomes transparent , if one starts 
with a Nambu-Goto-type Lagrangian ( the Jacobi's principle of least  action for 
a given E ) which is reparametrization invariant, 
\begin{eqnarray}
S &=& \int_{0}^{\tau} L d\tau = \int_{0}^{\tau} d\tau \sqrt{2m(E-V(r))(\frac{d\vec{x}}{d\tau})^{2}}\nonumber\\
&=& \int \sqrt{2m(E-V(r))(d\vec{x})^{2}}
\end{eqnarray}
The momenta conjugate to coordinates are  then defined by
\begin{equation}
\vec{p} = \frac{\partial L}{\partial \dot{\vec{x}}} = \sqrt{2m(E-V(r))}(\frac{d\vec{x}}{d\tau})/\sqrt{(\frac{d\vec{x}}{d\tau})^{2}}
\end{equation}
and  one obtains a vanishing Hamiltonian as a result of reparametrization
invariance and a first-class constraint $\phi$, which is the generator of 
reparametrization gauge symmetry,
\begin{eqnarray}
H &=& \vec{p}\dot{\vec{x}} - L = 0\nonumber\\
\phi (\vec{x}, \vec{p})& =& \frac{\vec{p}^{2}}{2m} +
 V(r) - E \simeq 0
\end{eqnarray}
Following Dirac[20], the total Hamiltonian is defined by
\begin{eqnarray}
H_{T}& =&  H + \alpha (\vec{x},\vec{p})\phi (\vec{x}, \vec{p})\nonumber\\
     & =&  \alpha (\vec{x},\vec{p})\phi (\vec{x}, \vec{p})\simeq 0
\end{eqnarray}
and the  function $\alpha (\vec{x}, \vec{p})$ specifies a choice of 
gauge and fixes the arbitrary parameter $\tau$ in (3.44), which parametrizes 
the orbit  for a given $E$. A change of the parameter $\tau$ to 
$\tau - \delta\beta (\tau, \vec{x}, \vec{p})$ is generated by 
$\delta\beta (\tau, \vec{x}, \vec{p})\alpha (\vec{x},\vec{p})\phi (\vec{x}, \vec{p}) = \delta\beta (\tau, \vec{x}, \vec{p})H_{T}$, for example,
\begin{eqnarray}
\delta\vec{x}(\tau) &=& \vec{x}^{\prime}(\tau) -  \vec{x}(\tau)\nonumber\\
&=& \vec{x}(\tau + \delta\beta) - \vec{x}(\tau)\nonumber\\
&=& \{ \vec{x}, \delta\beta H_{T}\}_{PB}\nonumber\\
&=& \delta\beta (\tau, \vec{x}, \vec{p})\frac{d}{d\tau}\vec{x}(\tau)
\end{eqnarray}
in terms of the Poisson bracket, since $\vec{x}^{\prime}(\tau - \delta\beta) =
\vec{x}(\tau)$.

Quantization is performed by
\begin{equation}
i\hbar\frac{\partial}{\partial\tau}\psi = \hat{H}_{T}\psi  
\end{equation}
with a physical state condition
\begin{equation}
\hat{\alpha}(\vec{x}, \vec{p})\hat{\phi} (\vec{x}, \vec{p})
\psi_{phy} = 0
\end{equation}
A specific choice of the gauge $\alpha (\vec{x}, \vec{p}) = r = \xi + \eta$ 
leads to the  Hamiltonian $\hat{H}_{T}$ in (3.7) and the choice $\alpha (\vec{x}, \vec{p}) = 1$ gives the original static Schroedinger equation(3.5), since the 
states $\psi$ in (3.7) and (3.5) are physical states. Eq.(3.49) gives rise to 
the evolution operator in (3.12), but the parameter $\tau$  need not be interpreted as a re-scaled time. In fact, the evolution operator (3.12), which essentially generates a gauge transformation,  deals with a 
sum over orbits in space instead of space-time. 

We now explain the relations (3.22)$\sim$(3.24) in a more concrete manner. We 
start with eq.(3.22) for a generic negative $E$
\begin{eqnarray}
G(E; \xi_{b},\eta_{b},\varphi_{b};\xi_{a},\eta_{a},\varphi_{a})
&=& \langle \xi_{b},\eta_{b},\varphi_{b}|\frac{\hbar}{
\hat{H}_{T}(E; \xi, \eta, \varphi)}|\xi_{a},\eta_{a},\varphi_{a}\rangle\nonumber\\
&\equiv& \sum_{n} \phi_{n}(E; \xi_{b},\eta_{b},\varphi_{b})\frac{\hbar}{\lambda_{n}(E)}\phi_{n}^{\star}(E; \xi_{a},\eta_{a},\varphi_{a})
\end{eqnarray}
with
\begin{eqnarray}
\hat{H}_{T}(E; \xi, \eta, \varphi)\phi_{n}(E; \xi,\eta,\varphi )
&=& \lambda_{n}(E)\phi_{n}(E; \xi,\eta,\varphi )\nonumber\\
\int \phi_{n}^{\star}(E; \xi,\eta,\varphi )\phi_{l}(E; \xi,\eta,\varphi ) dV_{0} &=& \delta_{n, l}\nonumber\\
\lambda_{n}(E) = 2n\hbar \sqrt{- \frac{E}{2m}} - e^{2}\nonumber\\
dV_{0} = 4\pi\times 2d\xi d\eta d\varphi
\end{eqnarray}
where we used the result in (3.34) and also the variables $(\xi, \eta, \varphi )$ instead of $(u, v, \varphi )$ for notational simplicity. The summation over $n$ in (3.51) is formal including the $n^{2}$ degeneracy. Note that the complete orthonormal states $\{ \phi_{n}\}$ in (3.52) are all {\em unphysical} off-
shell states. In path integral, the summation in (3.51) is exactly evaluated in
(3.20). 

We next rewrite $G(E; \xi_{b},\eta_{b},\varphi_{b};\xi_{a},\eta_{a},\varphi_{a})$ in terms of physical on-shell states by writing an unsubtracted dispersion
relation (i.e., paying attention only to the pole structure in $E$) as
\begin{equation}
G(E; \xi_{b},\eta_{b},\varphi_{b};\xi_{a},\eta_{a},\varphi_{a}) =
\sum_{n} \phi_{n}(E_{n}; \xi_{b},\eta_{b},\varphi_{b})\frac{\hbar}{(E_{n} - E)(- 
\frac{\partial \lambda_{n}(E_{n})}{\partial E_{n}})}\phi_{n}^{\star}(E_{n}; \xi_{a},\eta_{a},\varphi_{a})
\end{equation}
by noting 
\begin{equation}
\lambda_{n}(E) = \lambda_{n}(E_{n}) + (E-E_{n})\frac{\partial \lambda_{n}(E_{n})}{\partial E_{n}} = (E-E_{n})(\frac{-e^{2}}{2E_{n}})
\end{equation}
for $E \approx E_{n}$.

When one defines 
\begin{equation}
\psi_{n}(E_{n}; \xi,\eta,\varphi ) = \frac{1}{\sqrt{-\frac{\partial \lambda_{n}(E_{n})}{\partial E_{n}}}}\phi_{n}(E_{n}; \xi,\eta,\varphi )
\end{equation}
one can show the orthonormality relations
\begin{eqnarray}
\int \psi_{n}^{\star}(E_{n}; \xi,\eta,\varphi )\psi_{l}(E_{l}; \xi,\eta,\varphi )(\xi + \eta )dV_{0} &=&
\int \psi_{n}^{\star}(E_{n}; \xi,\eta,\varphi )\psi_{l}(E_{l}; \xi,\eta,\varphi )dV \nonumber\\
&=& \delta_{n,l}
\end{eqnarray}
with $dV = (\xi + \eta )dV_{0}$.
First of all, from the physical state condition 
\begin{eqnarray}
&&\hat{H}_{T}(E_{n}; \xi, \eta, \varphi )\phi_{n} (E_{n}; \xi, \eta, \varphi )
\\
&=&\{ \frac{1}{2m}(\hat{p}_{\xi} \xi \hat{p}_{\xi} + \hat{p}_{\eta}  \eta \hat{p}_{\eta}) + \frac{1}{8m}(\frac{1}{\xi} + \frac{1}{\eta})\hat{p}_{\varphi}^{2}  - E_{n}(\xi + \eta ) - e^{2} \}\phi_{n} (E_{n}; \xi, \eta, \varphi ) = 0\nonumber
\end{eqnarray}
one can establish the orthogonality relation
\begin{equation}
(E_{n} - E_{l})\int \phi_{n}^{\star}(E_{n}; \xi,\eta,\varphi )\phi_{l}(E_{l}; \xi,\eta,\varphi )(\xi + \eta )dV_{0} = 0
\end{equation}
for $n \neq l$. Also from the relation (3.57) and the fact that the ``eigenvalue'' $e^{2}$ is equally distributed for the kinetic and potential terms for
harmonic oscillators(in terms of $u$ and $v$ variables) , we have
\begin{equation}
- E_{n}\int \phi_{n}^{\star}(E_{n}; \xi,\eta,\varphi )\phi_{n}(E_{n}; \xi,\eta,\varphi )(\xi + \eta )dV_{0} = \frac{e^{2}}{2}
\end{equation}
namely
\begin{equation}
\frac{1}{-\frac{\partial \lambda_{n}(E_{n})}{\partial E_{n}}}\int \phi_{n}^{\star}(E_{n}; \xi,\eta,\varphi )\phi_{n}(E_{n}; \xi,\eta,\varphi )(\xi + \eta )dV_{0} = 1
\end{equation}
by noting (3.54). This proves (3.56).

From (3.53), we finally arrive at the expression
\begin{eqnarray}
G(E; \xi_{b},\eta_{b},\varphi_{b};\xi_{a},\eta_{a},\varphi_{a}) &=&
\sum_{n} \psi_{n}(E_{n}; \xi_{b},\eta_{b},\varphi_{b})\frac{\hbar}{E_{n} - E}
\psi_{n}^{\star}(E_{n}; \xi_{a},\eta_{a},\varphi_{a})\nonumber\\
&=& \langle \xi_{b},\eta_{b},\varphi_{b}|\frac{\hbar}{\hat{H}(\xi,\eta,\varphi ) - E}|\xi_{a},\eta_{a},\varphi_{a}\rangle \nonumber\\
&=& \frac{1}{4\pi}\langle \vec{x}_{b}|\frac{\hbar}{\hat{\vec{p}}^{2}/(2m)
- e^{2}/r - E}|\vec{x}_{a}\rangle 
\end{eqnarray}
which establishes the gauge independence of the Green's function for negative 
$E$. See also Ref.(14). Although we here used the same notation for the state $|\xi, \eta, \varphi
\rangle$ in (3.51) and (3.61), the meaning of these states are quite different.
This difference is explicitly exhibited in (3.22) $\sim$ (3.24). It is important to realize that the exact path integral is performed for the off-shell states in (3.51). As for the case of positive energy, one can define the Green's function
by analytic continuation in the starting expression in (3.51) and in the final 
expression in (3.61).

More generally, one can establish the gauge independence of the Green's function for an arbitrary choice of gauge condition $\alpha (\xi, \eta, \varphi ) =
f(\xi, \eta, \varphi )$ by noting
\begin{eqnarray}
\hat{H}_{T}(E; \xi, \eta, \varphi )\phi_{n}( E; \xi, \eta, \varphi ) &=&
[\hat{f}(\xi, \eta, \varphi )\hat{H}(\xi, \eta, \varphi ) - \hat{f}(\xi, \eta, \varphi ) E ]\phi_{n}( E; \xi, \eta, \varphi )\nonumber\\
&=& \lambda_{n}(E)\phi_{n} ( E; \xi, \eta, \varphi ) 
\end{eqnarray}
Namely
\begin{eqnarray}
\int \phi_{n}^{\star}( E; \xi, \eta, \varphi ) \hat{H}_{T}(E; \xi, \eta, \varphi )\phi_{n}( E; \xi, \eta, \varphi )dV_{f} &=& \lambda_{n}(E)\int \phi_{n}^{\star}( E; \xi, \eta, \varphi )\phi_{n}( E; \xi, \eta, \varphi )dV_{f}\nonumber\\
&=& \lambda_{n}(E)
\end{eqnarray}
with
\begin{equation}
dV_{f} = \frac{1}{f(\xi, \eta, \varphi )}dV
\end{equation}
which renders $\hat{H}_{T}(E; \xi, \eta, \varphi )$ defined by the gauge 
$\alpha = f(\xi, \eta, \varphi )$ hermitian. From (3.62) and (3.63), one derives
\begin{eqnarray}
\frac{\partial \lambda_{n}(E)}{\partial E}
&=& \int \phi_{n}^{\star}( E; \xi, \eta, \varphi )\frac{\partial}{\partial E} \hat{H}_{T}(E; \xi, \eta, \varphi )\phi_{n}( E; \xi, \eta, \varphi )dV_{f}\nonumber\\
&=& - \int \phi_{n}^{\star}( E; \xi, \eta, \varphi )\phi_{n}( E; \xi, \eta, \varphi )f(\xi, \eta, \varphi ) dV_{f}\nonumber\\ 
&=& - \int \phi_{n}^{\star}( E; \xi, \eta, \varphi )\phi_{n}( E; \xi, \eta, \varphi )dV
\end{eqnarray}
If one uses this relation for $E = E_{n}$ in (3.53), one arrive at the expression (3.61) starting with an arbitrary gauge condition $\alpha = f(\xi, \eta, \varphi )$ .

\section{A related topic : Stark effect}

 As an interesting implication of the present treatment of the separable Hamiltonian of Liouville-type, we comment on  the Stark effect (the hydrogen atom inside a
constant external electric field ${\cal E}$) described by a Hamiltonian
\begin{eqnarray}
H &=& \frac{1}{2m}\vec{p}^{2} - \frac{e^{2}}{ r} -e{\cal E} z\nonumber\\
  &=& \frac{1}{2m(\xi + \eta)}(\xi p_{\xi}^{2} + \eta p_{\eta}^{2}) +
\frac{1}{8m\xi\eta}{p_{\varphi}}^{2} - \frac{e^{2}}{\xi + \eta} - e{\cal E}(
\eta - \xi )
\end{eqnarray}
We thus analyze the total Hamiltonian defined by 
\begin{equation}
\hat{H}_{T} = \frac{1}{2m}\vec{p}_{u}^{2} + \frac{m\omega^{2}}{2}\vec{u}^{2} +
   \frac{1}{2m}\vec{p}_{v}^{2} + \frac{m\omega^{2}}{2}\vec{v}^{2} - e^{2}
   - \frac{1}{4}g(\vec{v}^{4} - \vec{u}^{4})
\end{equation}
with a constraint 
\begin{equation}
\hat{p}_{\varphi} - \hat{p}_{{\varphi}}^{\prime} = 0
\end{equation}
for the coodinates defined in (3.10). The quartic coupling constant $g$ is given by 
\begin{equation}
g = \frac{1}{4} e{\cal E}
\end{equation}
We can thus analyze the Stark effect on the basis of 
\begin{eqnarray}
&&\langle \vec{u}_{b}, \vec{v}_{b}| e^{ -i \hat{H}_{T} \tau/\hbar} | \vec{u}_{a}, \vec{v}_{a}\rangle\nonumber\\
&=& e^{ie^{2}\tau}
\langle \vec{u}_{b}| e^{-(i/\hbar) (\frac{1}{2m}\hat{\vec{p}}_{u}^{2} + \frac{m\omega^{2}}{2}\vec{u}^{2} + \frac{1}{4}g\vec{u}^{4}) \tau} | \vec{u}_{a}\rangle\nonumber\\
&&\times \langle \vec{v}_{b}| e^{-(i/\hbar) (\frac{1}{2m}\hat{\vec{p}}_{v}^{2} + \frac{m\omega^{2}}{2}\vec{v}^{2} - \frac{1}{4}g\vec{v}^{4}) \tau} | \vec{v}_{a}\rangle\nonumber\\
&=& e^{ie^{2}\tau}
\int {\cal D}\vec{p}_{u}{\cal D}\vec{u}\exp\{ (i/\hbar)\int_{0}^{\tau}[\vec{p}_{u}\dot{\vec{u}} - (\frac{1}{2m}\vec{p}_{u}^{2} + \frac{m\omega^{2}}{2}\vec{u}^{2} + \frac{1}{4}g\vec{u}^{4})] d\tau \}\nonumber\\
&&\times \int {\cal D}\vec{p}_{v}{\cal D}\vec{v}\exp\{ (i/\hbar)\int_{0}^{\tau}[\vec{p}_{v}\dot{\vec{v}} - (\frac{1}{2m}\vec{p}_{v}^{2} + \frac{m\omega^{2}}{2}\vec{v}^{2} - \frac{1}{4}g\vec{v}^{4})] d\tau \}
\end{eqnarray}

This problem in the conventional formulation has been recently analyzed by 
K. Hiraizumi, Y. Ohshima and H. Suzuki[22] as an application of the resummation 
technique of perturbation series for quantum tunneling [23], which was 
established for a system of 
anharmonic oscillators[24]. It is interesting that this problem is in fact {\em identical} to the (two-dimensional) anharmonic oscillator with an unstable quartic coupling in the 
present formulation.

\section{One-dimensional quantum gravity}

A way  alternative to (3.44)  to see the physical meaning of the parameter 
$\tau$ is to study the one-dimensional quantum gravity coupled to matter variables $\vec{x}$
defined by
\begin{equation}
\int \frac{{\cal D}\vec{x}{\cal D}h}{gauge\  volume} exp\{ i\int_{0}^{\tau} Lhd\tau\}
\end{equation}
with 
\begin{equation}
L = \frac{m}{2h^{2}}(\frac{d\vec{x}}{d\tau})^{2} - V(r) + E
\end{equation}
where $h$ stands for the einbein, a one-dimensional analogue of vierbein $h_{\mu}^{a}$, and $h= \sqrt{g}$ in one-dimension. In this Section, we set $\hbar = 1$.
If one uses the solution of the equation of motion for $h$ defined by the Lagrangian ${\cal L} = L h$, the action  in (5.1) is reduced to the one appearing in the Jacobi's principle of least action (3.44).
See Ref.[25] for a related problem in the context of a relativistic particle. The canonical Liouville measure is not reparametrization invariant in general[26], and it needs to be proved. This problem is analogous to the Polyakov-type path integral in string theory[27] - [29]. We here show that the naive canonical Liouville measure is gauge invariant up to a renormalization of the cosmological term ( or  
energy eigenvalue E). This renormalization is however  universal for any algebraic gauge fixing of the form
\begin{equation}
h(\tau) = f(\vec{x}(\tau))
\end{equation}
for a gauge fixing function $f(\vec{x}(\tau))$. 
The analysis presented below, which is formal but is known to work in string 
theory where no simple discretization is known, may be useful to understand certain  formal aspects such as gauge invariance in the point-particle path 
integral.

We here analyze the BRST invariant path integral for  (5.1) by using the Faddeev-Popov procedure, namely, we replace the naive measure in (5.1) by
\begin{eqnarray}
&&\int {\cal D}(\sqrt{h}\vec{x}){\cal D}\sqrt{h}{\cal D}(h^{3/2}c){\cal D}\bar{c}{\cal D}B \nonumber\\
&&\times exp\{i \int_{0}^{\tau}  L h d\tau + i\int_{0}^{\tau} [ B (\sqrt{h} - \sqrt{f}) - i \frac{1}{2}\bar{c}\sqrt{h}\partial_{\tau}c 
         - i \bar{c}c\partial_{\tau}(\sqrt{h} - \sqrt{f})] d\tau\}
\nonumber\\
&& \equiv \int d\mu \exp [ i\int_{0}^{\tau} {\cal L}_{eff} d\tau]
\end{eqnarray}
The BRST transformation is defined as a translation in the Grassmann parameter
, $\theta \rightarrow \theta + \lambda$, in the superfield notation(note that 
$\theta^{2} = \lambda^{2} = \theta\lambda + \lambda\theta = \theta c(\tau) +
c(\tau)\theta = 0$)
\begin{eqnarray}
\vec{x}(\tau, \theta) &=& \vec{x}(\tau) + i \theta c(\tau)\partial_{\tau}\vec{x}(\tau)\nonumber\\
\sqrt{h(\tau, \theta)} &=& \sqrt{h(\tau)} + 
i\theta [c(\tau)\partial_{\tau} + \frac{1}{2}(\partial_{\tau}c(\tau))]\sqrt{h(\tau)}\nonumber\\
\sqrt{h}\vec{x}(\tau, \theta) &=& \sqrt{h}\vec{x}(\tau) + i \theta [ c(\tau)\partial_{\tau} + \frac{1}{2}(\partial_{\tau}c(\tau))]\sqrt{h}\vec{x}(\tau)\nonumber\\
\sqrt{f(\vec{x}(\tau, \theta))} &=& \sqrt{f(\vec{x}(\tau))} + i \theta c(\tau)
\partial_{\tau} \sqrt{f(\vec{x}(\tau))}\nonumber\\
c(\tau, \theta) &=& c(\tau) + i\theta c(\tau)\partial_{\tau}c(\tau)\nonumber\\
\bar{c}(\tau, \theta) &=& \bar{c}(\tau) + \theta B(\tau)
\end{eqnarray}
For example, the BRST transformation is given by 
\begin{eqnarray}
\delta \sqrt{h(\tau)} &=& i\lambda [c(\tau)\partial_{\tau} + \frac{1}{2}(\partial_{\tau}c(\tau))]\sqrt{h(\tau)}\nonumber\\
\delta (\sqrt{h}\vec{x}(\tau)) &=& i\lambda [c(\tau)\partial_{\tau} + \frac{1}{2}(\partial_{\tau}c(\tau))]\sqrt{h}\vec{x}(\tau)\nonumber\\
\delta (h^{3/2}dc(\tau)) &=& i\lambda [c(\tau)\partial_{\tau} + \frac{1}{2}(\partial_{\tau}c(\tau))](h^{3/2}dc(\tau))
\end{eqnarray}
In the last relation for the ghost variable $c(\tau)$, we consider the differential $dc(\tau)$:  one may write $h^{3/2}dc =  d(h^{3/2}c)$ for a fixed metric $h(\tau)$, which is the case 
required to study the BRST invariance of the path integral measure in (5.4).   Note that all the variables in (5.6) have the same BRST transformation law which is anomaly free[26]; the combination ${\cal D}\bar{c}{\cal D}B$ is also 
manifestly BRST invariant. These properties  in turn ensure the BRST
invariance of the path integral measure in (5.4). One can also confirm that the
action in (5.4) is also BRST invariant.  

By using the BRST invariance of the action and the measure in (5.4), one can 
prove that the path integral (5.4) is independent of the choice of $f(\vec{x}(\tau))$ as follows: Under an infinitesimal change of $f(\vec{x}(\tau))$, the path integral (5.4) changes as 
\begin{eqnarray}
&&- i \int_{0}^{\tau}d\tau \langle [B(\tau)\delta\sqrt{f(\tau)} - i 
\bar{c}(\tau)c(\tau)\partial_{\tau}\delta\sqrt{f(\tau)}]\rangle\nonumber\\
&&= - i\int d\mu [B(\tau)\delta\sqrt{f(\tau)} - i 
\bar{c}(\tau)c(\tau)\partial_{\tau}\delta\sqrt{f(\tau)}] 
e^{ i\int_{0}^{\tau} {\cal L}_{eff} d\tau}
\end{eqnarray}
where $\langle B\delta\sqrt{f} -i \bar{c}c\partial_{\tau}\delta\sqrt{f}\rangle$ denotes the averaging in  the path integral (5.4). Next we note the BRST identity
\begin{eqnarray}
&&\langle \bar{c}(\tau)\delta\sqrt{f(\vec{x}(\tau))} \rangle \nonumber\\
&&= \int d\mu \{\bar{c}(\tau)\delta\sqrt{f(\vec{x}(\tau))}\}  
e^{ i\int_{0}^{\tau} {\cal L}_{eff} d\tau}\nonumber\\
&&= \int d\mu^{\prime} \{\bar{c}(\tau)^{\prime}\delta\sqrt{f(\vec{x}(\tau))^{\prime}}\}  
e^{ i\int_{0}^{\tau}{{\cal L}_{eff}}^{\prime} d\tau}\nonumber\\
&&= \int d\mu \{\bar{c}(\tau)\delta\sqrt{f(\vec{x}(\tau))} + \lambda [B(\tau)\delta\sqrt{f(\tau)} - i 
\bar{c}(\tau)c(\tau)\partial_{\tau}\delta\sqrt{f(\tau)}]\}  
e^{ i\int_{0}^{\tau}{\cal L}_{eff} d\tau}\nonumber\\
&&= \langle \bar{c}(\tau)\delta\sqrt{f(\vec{x}(\tau))} \rangle + \lambda\langle B(\tau)\delta\sqrt{f(\tau)} - i 
\bar{c}(\tau)c(\tau)\partial_{\tau}\delta\sqrt{f(\tau)}\rangle
\end{eqnarray}
where the primed variables stand for the BRST transformed variables such as
$  \sqrt{h(\tau)^{\prime}} = \sqrt{h(\tau)} + i\lambda [c(\tau)\partial_{\tau} + \frac{1}{2}(\partial_{\tau}c(\tau))]\sqrt{h(\tau)}$, $\delta\sqrt{f(\vec{x}(\tau))^{\prime}} = \delta\sqrt{f(\vec{x}(\tau))} + i\lambda c(\tau)\partial_{\tau}\delta\sqrt{f(\vec{x}(\tau))}$  and ${\bar{c}(\tau)}^{\prime} = \bar{c}(\tau) + \lambda B(\tau)$.
 The first equality in (5.8) holds since the path integral is independent of the naming of path integration variables, and the second equality holds because of the BRST invariance of the measure $d\mu^{\prime} = d\mu$ and the effective action ${{\cal L}_{eff}}^{\prime} = {\cal L}_{eff}$.
 (5.8) shows that 
$\langle B\delta\sqrt{f} -i \bar{c}c\partial_{\tau}\delta\sqrt{f} \rangle = 0$, and thus (5.7) vanishes. Namely, the path integral (5.4) is independent of the choice of $f(\vec{x}(\tau))$, provided that one specifies the gauge invariant  initial and final states. 

The path integral (5.4) is rewritten as 
\begin{equation}
\int {\cal D}\tilde{\vec{p}}{\cal D}\tilde{\vec{x}}{\cal D}\sqrt{h}
{\cal D}\tilde{c}{\cal D}\bar{c}{\cal D}B 
exp\{ i\int_{0}^{\tau} d\tau [( \vec{p}\dot{\vec{x}} - Hh) +
B(\sqrt{h} - \sqrt{f}) - i\frac{1}{2}\bar{c}\sqrt{h}\partial_{\tau}
( \frac{1}{h^{3/2}}\tilde{c})]\}
\end{equation} 
with 
\begin{eqnarray}
H &=& \frac{1}{2m}\vec{p}^{2} + V(r) - E \nonumber\\
\tilde{\vec{p}} &=& \sqrt{h}\vec{p}\nonumber\\
\tilde{\vec{x}} &=& \sqrt{h}\vec{x}\nonumber\\
\tilde{c} &=& h^{3/2}c
\end{eqnarray}
In fact, after the path integral over $\tilde{\vec{p}}$, one recovers (5.4). 
We also set $\sqrt{h} - \sqrt{f} = 0$ after a partial integration in the ghost 
sector in (5.9). Note that the path integral
\begin{equation}
\int {\cal D}\tilde{\vec{p}}\exp\{  i\int [- \frac{h}{2m}{\vec{p}}^{2} 
+ \vec{p}\dot{\vec{x}} - \frac{m}{2h}{\dot{\vec{x}}}^{2}] d\tau\}
= \int {\cal D}\tilde{\vec{p}}\exp [ i\int \frac{-1}{2m}(\tilde{\vec{p}})^{2} d\tau]
\end{equation}
is a constant independent of the metric $h(\tau)$: This is another way to see why the variables with weight $1/2$ such as $\tilde{\vec{p}}$ and $\tilde{\vec{x}}$ are chosen for the world 
scalar quantities $\vec{p}$ and $\vec{x}$  as reparametrization invariant path integral variables.

In the above path integral, it is important to recognize that the singular (time-derivative) terms in the Lagrangian (5.9) are written as 
\begin{equation}
\int {\cal D}\tilde{\vec{p}}{\cal D}\tilde{\vec{x}}
{\cal D}\tilde{c}{\cal D}\bar{c} \exp\{  \int_{0}^{\tau} d\tau [ i\tilde{\vec{p}}\frac{1}{\sqrt{h}}\partial_{\tau}( \frac{1}{\sqrt{h}}\tilde{\vec{x}}) + \frac{1}{2}\bar{c}\sqrt{h}\partial_{\tau}
( \frac{1}{h^{3/2}}\tilde{c})]\}
\end{equation}
Those singular terms have a Weyl invariant structure; namely, the $h$-dependence 
can be completely removed by a suitable scale transformation of $\tilde{\vec{p}},\tilde{\vec{x}},\tilde{c}$ and $\bar{c}$ such as $ \tilde{\vec{x}}\rightarrow \sqrt{h}\tilde{\vec{x}}$ and $\bar{c}\rightarrow (1/\sqrt{h})\bar{c}$. In this process of scaling, one obtains a Jacobian (or anomaly), which can be integrated to a Wess-Zumino term. At the same time, the path integral variables $\tilde{\vec{p}}$ and $\tilde{\vec{x}}$ are reduced to the naive ones. In the present case, one can confirm that the Jacobian (Weyl anomaly) has a form
\begin{equation}
M \int  h\delta\beta (\tau)d\tau
\end{equation}
for an infinitesimal scale transformation
\begin{eqnarray}
\bar{c}(\tau)&\rightarrow&  e^{- \delta\beta (\tau)}\bar{c}(\tau)\nonumber\\
\tilde{c}(\tau)&\rightarrow&  e^{3 \delta\beta (\tau)}\tilde{c}(\tau)\nonumber\\
\tilde{\vec{p}}(\tau)&\rightarrow&  e^{\delta\beta (\tau)}\tilde{\vec{p}}(\tau)\nonumber\\
\tilde{\vec{x}}(\tau)&\rightarrow&  e^{\delta\beta (\tau)}\tilde{\vec{x}}(\tau)
\end{eqnarray}
This evaluation of the Jacobian is performed for $\tilde{\vec{p}}$ and $\tilde{\vec{x}}$, for example, by[28][30]
\begin{eqnarray}
&&\lim_{M \rightarrow \infty}\int\frac{dk}{2\pi}e^{-ik\tau}\exp\{(\frac{1}{\sqrt{h}}\partial_{\tau}\frac{1}{\sqrt{h}})^{\dagger}(\frac{1}{\sqrt{h}}\partial_{\tau}\frac{1}{\sqrt{h}})/M^{2}\}e^{ik\tau}\nonumber\\
&&= \lim_{M \rightarrow \infty}M\int\frac{dk}{2\pi}\exp\{[\frac{1}{\sqrt{h}}(\partial_{\tau}/M + ik)\frac{1}{\sqrt{h}}][\frac{1}{\sqrt{h}}(\partial_{\tau}/M + ik)\frac{1}{\sqrt{h}}]\}\nonumber\\
&& = \lim_{M \rightarrow \infty}M\int\frac{dk}{2\pi}\exp\{ -k^{2}/h^{2}\}( 1 + 
O(\frac{1}{M^{2}}))\nonumber\\ 
&& = \frac{1}{2\sqrt{\pi}}M h \ \ \ \ for \ \ M\rightarrow \infty
\end{eqnarray}
which gives a term of a general structure as in (5.13).
The anomaly calculation is specified by the basic operators appearing in (5.12) 
\begin{equation}
\frac{1}{\sqrt{h}}\partial_{\tau}\frac{1}{\sqrt{h}}\ \ \ , or \ \ 
\sqrt{h}\partial_{\tau}\frac{1}{h^{3/2}}
\end{equation}
both of which give the same form of anomaly proportional to $h$,  as in (5.15).
See also Ref.[28]. The overall sign of the Jacobian is specified depending on the statistics of 
each variable, i.e., a Grassmann variable gives an extra minus sign. Since only the most singular term survives in (5.15), one can confirm that the knowledge of the most singular terms in the Lagrangian (5.9), i.e.,the time-derivative terms in  (5.12), is sufficient to 
calculate anomaly. If one denotes the integrated anomaly ( Wess-Zumino term) by $\Gamma (h)$, we  have from (5.13) 
\begin{equation}
\Gamma (h) - \Gamma (he^{-2\delta\beta}) = \int 2\delta\beta (\tau)h(\tau)\frac{\partial\Gamma}{\partial h(\tau)}d\tau = M\int h(\tau)\delta\beta (\tau)d\tau
\end{equation}
since $h$ is transformed to $he^{-2\delta\beta}$ by (5.14), and  we obtain
\begin{equation}
\Gamma (h) = M\int h(\tau)d\tau
\end{equation}
with  a suitable (infinite) number $M$. This $\Gamma (h)$  is added to the action in (5.9) and it has a form of the cosmological (or energy) term in (5.9).

This calculation of the anomaly is quite general: In two-dimensions, the Weyl anomaly has a structure $M^{2}\sqrt{g} + \sqrt{g}R$,[27] - [29]. In one-dimension, the curvature term containing $R$ does not exist, and only the cosmological term arises if one performs a reparametrization invariant calculation. This Weyl anomaly  renormalizes the bare cosmological term in (5.9),
\begin{equation}
E + M = E_{r}
\end{equation}
with $E_{r}$ a renormalized energy parameter.
But this renormalization is independent of the choice of the gauge fixing function $f(\vec{x}(\tau))$ in (5.4).

We cannot assign a physical significance  to the precise value of the renormalization in (5.19), since the linear divergence in (5.15) is regularization dependent. Only the Weyl invariant structure in (5.12) for the choice of BRST invariant measure and the extraction of the metric dependence from the singular terms in (5.12) as a Wess-Zumino term have a well-defined physical meaning.

After this procedure of eliminating the $h$-dependence from  the time-derivative 
terms  and then the integration over $B$ and $h$, one obtains  a  path integral
\begin{equation}
\int {\cal D}\vec{p}{\cal D}\vec{x}
{\cal D}c{\cal D}\bar{c} 
exp\{ i\int_{0}^{\tau} d\tau [ \vec{p}\dot{\vec{x}} - f(\vec{x}(\tau))H  -  i\frac{1}{2}\bar{c}\partial_{\tau}c]\} 
\end{equation}
One may ignore the {\em decoupled} ghost sector, which is first order in $\partial_{\tau}$ in the present case and does not contribute to the Hamiltonian.
One thus  finally arrives at the phase space path integral  
\begin{equation}
\int {\cal D}\vec{p}{\cal D}\vec{x} 
exp\{ i\int_{0}^{\tau_{f}} d\tau [ \vec{p}\dot{\vec{x}} - f(\vec{x}(\tau))H ]\} 
\end{equation}
where  the overall normalization factor, which is independent of $f(\vec{x})$
but can depend on $\tau_{f}$, is fixed to be the same as for $f(\vec{x}) = 1$ in (5.21).

When one cuts the interval $[0, \tau_{f}]$ into meshes, the path integral measure in (5.21) is defined by
\begin{eqnarray}
dV_{f}(\vec{x}) &=& \frac{1}{f(\vec{x})}d^{3}x \nonumber\\
dV_{f}(\vec{p}) &=& f(\vec{x})d^{3}p
\end{eqnarray}
for each mesh point, and one performs integral over the momentum first. Since 
one has one extra momentum integration relative to the coordinate one for fixed  end points $\vec{x}_{a}(0)$ and $\vec{x}_{b}(\tau_{f})$, the external states 
are also specified for the volume element $dV_{f}(\vec{x})$.

A way to provide gauge invariant external states is to consider a trace by integrating over $\vec{x}_{a}(0)= \vec{x}_{b}(\tau_{f})$ in (5.21), and let $E_{r} 
\rightarrow E_{n}$. In this limit the path integral is reduced to 
\begin{equation}
\sum \int dV_{f}(\vec{x}_{a})\phi_{n}(E_{n};\vec{x}_{a})\phi_{n}^{\star}(E_{n};\vec{x}_{a}) = n^{2}
\end{equation}
where we used the notation in (3.57), and the summation is over the $n^{2}$
degenerate states. The states $\phi_{n}(E_{n};\vec{x})$ form a complete physical set for $\hat{H}_{T}(E_{r}=E_{n})$.

A less trivial way to incorporate physical external states is to  consider the Green's function
\begin{equation}
G(E_{r};\vec{x}_{b},\vec{x}_{a}) = i\int_{0}^{\infty} d\tau_{f} \int {\cal D}\vec{p}{\cal D}\vec{x} 
exp\{ i\int_{0}^{\tau_{f}} d\tau [ \vec{p}\dot{\vec{x}} - f(\vec{x}(\tau))H ]\}
\end{equation}
and look at the pole position at $E_{r} \simeq E_{n}$. 
In this case, one has (see also eq(3.53) )
\begin{eqnarray}
\int dV_{f}(\vec{x}_{b})dV_{f}(\vec{x}_{a})\phi_{n}^{\star}(E_{n};\vec{x}_{b})
G(E_{r};\vec{x}_{b},\vec{x}_{a})\phi_{n}(E_{n};\vec{x}_{a})&=& \frac{1}{\lambda_{n}(E_{r})}\nonumber\\
&=& \frac{1}{(E_{r}-E_{n})}\frac{1}{(-\frac{d\lambda_{n}(E_{n})}{dE_{n}})}
\end{eqnarray}
If one recalls the definition of $\psi_{n}(E_{n};\vec{x})$ in (3.55), one 
concludes(see eq.(3.61))
\begin{equation}
G(E_{r};\vec{x}_{b},\vec{x}_{a}) = \sum_{n} \psi_{n}(E_{n};\vec{x}_{b})\frac{1}{E_{r}-E_{n}}\psi_{n}^{\star}(E_{n};\vec{x}_{a}) 
\end{equation}
in the sense of an unsubtracted dispersion relation. The pole positions and 
the residue functions in this relation are all gauge independent.

In the form of the quantity $\langle n|\frac{1}{\hat{H}_{T}}|n \rangle$ in (5.25), one needs to adjust the ``wave function renormalization
factor'' $\sqrt{-\frac{d\lambda_{n}(E_{n})}{dE_{n}}}$ for each external state 
to obtain a gauge independent physical quantity, but in the form $G(E_{r};\vec{x}_{b},\vec{x}_{a})$ in (5.24) one need not supply the renormalization factor.

An interesting implication of the path integral approach(5.4)  is that one may understand the $\tau$-integral in (5.24)
as an integral over the deformation parameter ( or moduli) of an analogue of the world-line.  In the present path integral, one sums over all the possible deformation of the ``world-line'' as well as the paths in space to obtain a physically meaningful 
geometrical quantity. Another interpretation of $\tau$ may be to regard it as 
an analogue of a proper time[31].

\section{ Conclusion}
An attempt to solve the Green's function for the hydrogen atom exactly in the 
path integral[4] opened a 
new avenue for the path integral treatment of a general separable Hamiltonian 
of Liouville-type.  This  new view point,
which has been shown to be  based on the Jacobi's principle of least action ,  provides a more flexible framework of  path integral to deal with a wider class of problems of physical interest. 
The Jacobi's principle of least action , besides being reparametrization 
invariant, gives  an attractive geometrical picture of particle orbits in a 
curved space deformed  by the potential. On the other hand, the fundamental 
space-time picture of the conventional Feynman path integral, which is associated with the Hamilton's principle of stationary action, is lost. ( A 4-dimensional picture is however recovered by a generalization of the Jacobi's principle 
for a relativistic particle, $S = -m\int d\tau \sqrt{(dx^{\mu}/d\tau)^{2}}$, and an arbitrary parameter $\tau$ is identified with a proper time in quantum 
theory[31]).

In the present paper, we discussed some of the basic issues 
related to this new approach to  the path integral from a view point of 
general gauge theory.
We have shown the gauge 
independence of the Green's function. A BRST analysis of the problem from a 
view point of one-dimensional quantum gravity is presented.
We also commented on a 
possible application of this  scheme to the Stark effect and a resummation of perturbation series.
\

I thank C. Bernido for calling the path integral of the hydrogen atom to my 
attention.\\

[{\bf Note added}]\\
After submitting the present paper for publication, the papers quoted in 
Ref.[32] came to my attention. These papers deal with some related matters 
of the path integral treatment of the hydrogen atom.

\end{document}